\documentclass[aps,prb,twocolumn,groupedaddress,showpacs]{revtex4}
\usepackage[english]{babel}
\usepackage{graphicx}
\usepackage{amssymb}
\usepackage{amsmath}

\begin{document}

\title{Vortex fluctuations and freezing of dipolar-coupled granular moments in thin ferromagnetic films}

\author{J\"urgen K\"otzler}
\affiliation{Institut f\"ur Angewandte Physik und Zentrum f\"ur Mikrostrukturforschung, Universit\"at Hamburg, Jungiusstrasse 11, D-20355 Hamburg, Germany}
\author{Detlef G\"orlitz}
\affiliation{Institut f\"ur Angewandte Physik und Zentrum f\"ur Mikrostrukturforschung, Universit\"at Hamburg, Jungiusstrasse 11, D-20355 Hamburg, Germany}
\author{Malte Kurfi\ss}
\affiliation{Institut f\"ur Angewandte Physik und Zentrum f\"ur Mikrostrukturforschung, Universit\"at Hamburg, Jungiusstrasse 11, D-20355 Hamburg, Germany}
\author{Lars von Sawilski}
\affiliation{Institut f\"ur Angewandte Physik und Zentrum f\"ur Mikrostrukturforschung, Universit\"at Hamburg, Jungiusstrasse 11, D-20355 Hamburg, Germany}
\author{Elena Y. Vedmedenko}
\affiliation{Institut f\"ur Angewandte Physik und Zentrum f\"ur Mikrostrukturforschung, Universit\"at Hamburg, Jungiusstrasse 11, D-20355 Hamburg, Germany}

\date{\today}

\begin{abstract}
Below the Curie temperature $T_c$ of a Heusler-alloy film, consisting of densely packed, but exchange-decoupled nanograins, the spontaneous magnetization $M_s(T)$ and static in-plane susceptibility $\chi_{\|}(T)$ increase very slowly signalizing a suppression of magnetization fluctuations. The unpredicted variation $\chi_{\|}(T)\sim G_d^{2}(T)$, where $G_d\sim M_s^2$ is the intergranular dipolar coupling, and also the magnetic freezing observed in the dynamic susceptibility at lower temperatures is quantitatively reproduced by Monte Carlo (MC) simulations with 10$^{4}$ dipolar-coupled moments on a disordered triangular lattice. At high temperatures, the MC spin configurations clearly reveal a dense gas of local vortex structures, which at low temperatures condense in regions with stronger disorder. This vortex depletion upon decreasing temperature seems to be responsible for the observed \textit{increase} of the magnetic relaxation time. For weak disorder, the temperature dependence of the MC vorticity and a singularity of the specific heat at $T_v=\frac{1}{2}G_d(T_v)/k_B$ indicate a thermal transition from a vortex gas to a state with a single vortex center plus linear vortex structures.
\end{abstract}


\maketitle

\section{Introduction}
Thin ferromagnetic films are not only a very popular playground for physical research, but they also serve as the basic building blocks of the information technology (see e.g. Ref.\onlinecite{HB05}). The desire for high-density storage media has motivated numerous studies on the preparation of nanomagnetic arrays and on the effects of the dipolar interaction between the nanomagnets on their dynamic behavior. In diluted arrays, the superparamagnetic fluctuations across the Ne\'el \cite{Nee49} energy barrier, $E_b = K_A V_g$, depending on the anisotropy $K_A$ and the volume of the nanograin $V_g$, limit the stability of the nanomagnetic moments. Upon increasing density of the moments, the dipolar coupling between the grains, $E_d$, gains importance and, due to its long range and angular anisotropy, the influence on the fluctuations and the dynamics of the nanomoments is extremely complex. This issue is of current interest and being studied from various directions.

Recent theoretical work on the effect of structural disorder on the dipolar interaction in 2D ensembles of nanoparticles lead to the conclusion that the average dipole energy decreases with increasing structural disorder \cite{JP03,Jen06}. On square/triangular periodic lattices \cite{PP02}, the disorder destroys the antiferromagnetic/ferromagnetic dipolar ground state in favor of complicated non-collinear structures \cite{Pet06} with complex energy landscape \cite{GPR00} leading to the spin-glass like behavior. In a strongly coupled system with a large number of metastable minima an angular rotation of a macroscopic magnetic moment of a given particle should cause neighboring particles to vary their magnetic directions as well, {\emph i.e.} the cooperative reversal occurs \cite{Jen06,CJR04}. While the static distribution of the energy barriers has been discussed in the literature \cite{ADJ97, KT00, IL04}, the description of the dynamic properties of a 2D interacting system is still lacking. Here, we mention the dependence of the freezing temperature on disorder, which is directly related to the temperature fluctuations. Moreover, critical exponents in dipolar arrays and the dynamic susceptibility have not been systematically studied. Consequently, it is yet not known whether in the presence of disorder the magnetic fluctuations are dominated by vortex-type excitations or via collinear reorientation of the magnetic moments.

Upon increasing the coverage of a random assembly of nanometer-sized Fe-islands on $CaFe_2/Si(111)$ wafers, Scheinfein et al.\cite{SSH96} observed a change from superparamagnetism to in-plane ferromagnetism. Based on MC simulations, they attributed this long-range ordering to an interparticle exchange interaction, noting that in the absence of exchange such transition would be destroyed by thermal fluctuations. Much attention focused on the hysteresis of the magnetization loops of two-dimensional (2D) arrays of nanomagnets, in particular on the interplay between different kinds of particle anisotropy, disorder, and the dipolar interaction. Hysteresis loops and coercive fields of 2D arrays of diluted Co nanomagnets \cite{RPL00}, of close-packed Co \cite{AML98,CJR04}, and of permalloy \cite{CJR04} particles were found to be in at least qualitative agreement with numerical evaluations of suitably chosen models for structure and interactions. More recent numerical work considered smaller arrays of dipolar coupled particles with anisotropy on 2D lattices to obtain spin configurations, which display hysteresis loops with a multistep structure \cite{KS04}. By ac susceptibility measurements in zero magnetic field, dipolar effects on the thermally activated magnetization dynamics have been detected on diluted hexagonal arrays of $Co$ nanoclusters \cite{LPT02} and on close-packed, i.e. strongly interacting iron oxide nanoparticles \cite{TM05}. In the first case, the slight enhancement of the relaxation time $\tau$ and the energy barrier $E_b$ was found to be consistent with a model of independent anisotropic grains with weak dipolar interaction. In the second case, stronger deviations from the Ne\'el-behavior have been interpreted in terms of a Vogel-Fulcher law for $\tau(T)$, i.e. spin-glass like collective dynamics \cite{TM05}.

Our work is devoted to the zero-field magnetization dynamics for a rather simple case that has not been persued in too great detail. We study a close-packed 2D array of magnetic moments, in which the dipolar coupling $G_d$ between the moments is much larger than any other energy, in particular, than volume, surface , and interface anisotropies. Experimentally, such system is realized by a granular thin film, in which the grain boundaries do not transmit the ferromagnetic exchange. 
We investigate a thin film of the ferromagnetic cubic Heusler alloy NiMnIn, where the exchange between the magnetic Mn-moments arises from delocalized 3d-electrons provided by the almost nonmagnetic Ni \cite{SSB04}, which are also responsible for the potential of these alloys as spin-injectors to lattice-matched semiconductors \cite{HLK04,DLX01,KKR04}. Another nice feature of the $Mn$-based alloy is the small anisotropy field ($H_a \leq 300~Oe$)\cite{AMA01}, most likely due to the weak coupling of the Mn-ion to the lattice. The large granular moment, $\mu_g(T=0)=1.1\cdot 10^6 \mu_B$, implies strong dipolar coupling, so that all experimental phenomena, in particular the magnetization dynamics are strongly collective. As a unique feature of the present films, we note that, by a variation of the temperature $G_d(T)$ can be decreased from the maximum value $G_d(0)/k_B=3\cdot10^4$~K to zero at $T_C$.  A deeper understanding of the dipolar-dominated magnetic fluctuations and of their dynamics is intended by means of MC simulations \cite{VGL99} of a closely related disordered system.

The outline of this work is the following. In Section II the results of the magnetic measurements and of a structural characterization of the granular film are described. The main emphasis is devoted to the temperature variation of the granular magnetic moment $\mu_g(T)$ and to the dc- and ac-susceptibilities in zero magnetic field, which probe the statics and dynamics of the magnetic fluctuations. By starting at $T_C$, the magnitude of the dipolar coupling, $G_d(T) \sim \mu_g^2(T)$, is increased up to large values to study its effect on the fluctuations. In Section III the MC simulations as a function of the normalized temperature $\widetilde{T}=k_BT/G_d$ are presented. The results for the static and the time dependent susceptibility for different degrees of disorder are directly compared to the experimental data. Based on the surprising good agreement with respect to temperature variation \underline{and} disorder, we then discuss the largely unexpected experimental findings in terms of the spin configurations provided by the MC simulations. Section IV contains the conclusions and some open questions.

\begin{figure}
\begin{center}
\includegraphics[width=8cm]{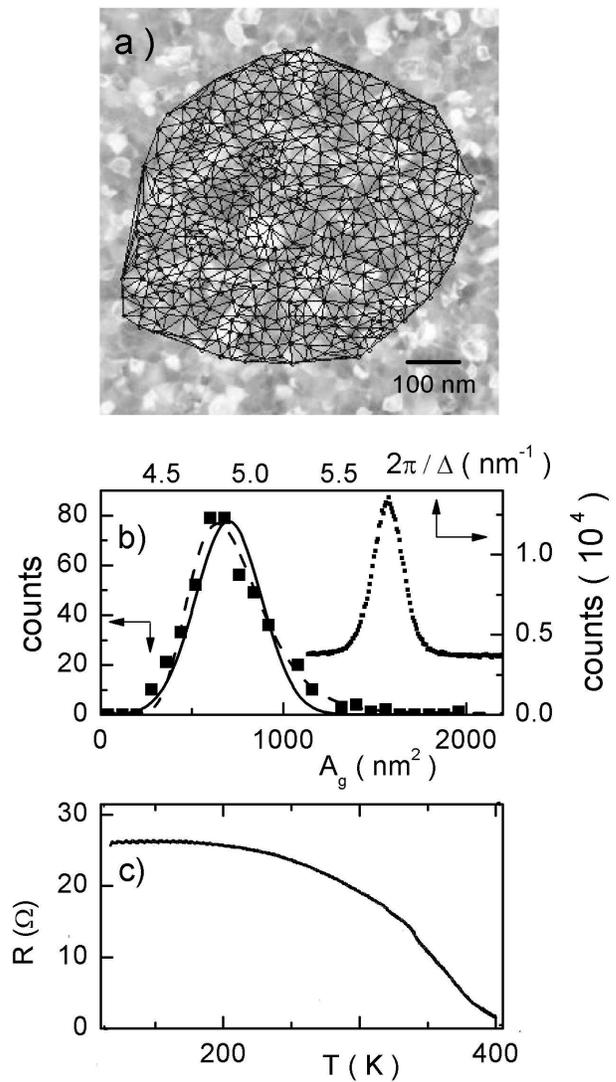}
\caption{\protect{(a) TEM image of the NiMnIn film with triangulation of the grain centers. (b) Distribution of grain areas obtained from the triangulation in (a) and of the grain boundary thicknesses (upper and right-hand scales, explained in the text). Solid curves are Gaussians, the dashed curve is a log-normal distribution with the same mean area $\overline{A_g}=680~nm^2$. (c) Temperature dependence of the resistance of the NiMnIn film on p-Si (from Ref.\onlinecite{KS05}).}}
\end{center}
\end{figure}

\section{Experimental Results}
The present film has been deposited on p-Si(100) by evaporating Ni and the stoichiometric MnIn alloy from two independent sources as described  in more detail elsewhere \cite{KSA05}. X-ray spectroscopy and diffraction revealed the so-called half-Heusler alloy NiMnIn, which can be described as the $L2_1$ structure of the full-Heusler alloy Ni$_2$MnIn (see e.g. Ref. \onlinecite{Web69}), where one of the four intervening fcc lattices for each metal ion is missing. The granular structure of the film has been evidenced by atomic force microscopy (AFM) and also by transmisssion electron microscopy (TEM) on a film simultaneously prepared on amorphous carbon, see Fig.1(a). AFM scans displayed very sharp film edges and provided for the thickness t=35$\pm$1~nm. We analyzed the TEM image by means of a  triangulation \cite{tri}: taking the grain centers , i.e. the vertices shown in Fig.1(a), as lattice points, one finds the distribution of the intergranular distances $d$, which define the areas of the hexagonal grains, $A=d^2\sqrt{3}/2$. The results depicted in Fig.1(b) are consistent with both the gaussian and the log-normal distributions and the same mean grain area $A_g= 680\pm40$~nm$^2$. This implies for the mean distance between the grain centers $d_g=28.0$~nm. We will refer to these results below, when analyzing th ac-susceptibilities.

\begin{figure}
\begin{center}
\includegraphics[width=8cm]{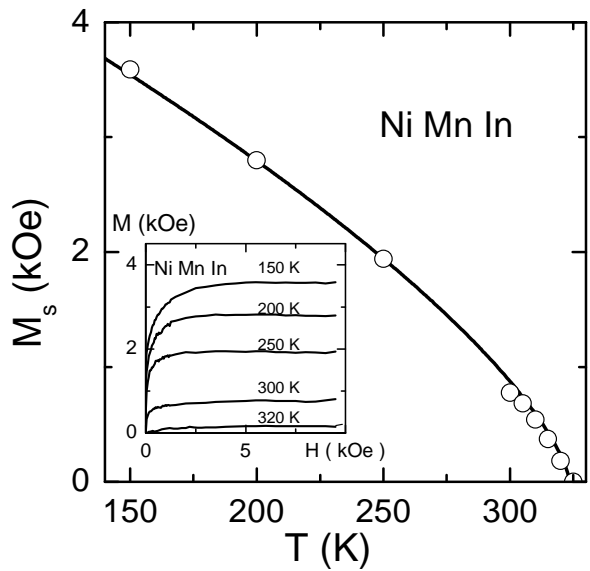}
\caption{Temperature dependence of the technically saturated magnetization $M_s$ of a NiMnIn Heusler-alloy film. $M_s(T)$ is defined by the plateaus of the magnetization isotherms below the Curie-temperature, shown by the inset. The solid curve fits the data to the power law, $M_s\sim(1-T/T_C)^{\beta}$, with the large exponent $\beta$=0.66.}
\end{center}
\end{figure}

Using a digital processing of the TEM image, the contrasts produced by the grain boundaries have been determined and analyzed by a Fast Fourier Transformation \cite{FFT}. The results shown in Fig.1(b), display a rather sharp Gaussian peak at the spatial frequency $2\pi/\Delta=5.77~nm^{-1}$, which implies for the mean grain boundary thickness $\Delta=(1.1\pm0.1)nm$. We believe that this rather large value for $\Delta$ is responsible for the negligible contribution of the Heusler film to the resistance curve R(T) presented in Fig.1(c): the plateau of R at low temperatures and the drop near room temperature arise fom a highly doped $p^+-Si$ layer and electron-hole excitations in the $Si$-wafer, respectively \cite{KSA05}. The accuracy of the data sets a lower (nonmetallic) limit of $\rho \gtrsim 1~m\Omega cm$ to the Heusler-film. As the ferromagnetic exchange between localized $Mn$-moments in the cubic structure results from the conduction electrons\cite{SSB04}, such coupling across the wide and disordered boundaries appears to be rather unlikely.

The magnetic measurements on the NiMnIn film (11~mm$^2$ area) have been performed by a SQUID magnetometer (MPMS2, Quantum Design, San Diego). Immediately below the Curie temperature of NiMnIn, the magnetization isotherms rise rather steeply to their technical saturation value $M_s(T)$ (see inset to Fig. 2). According to Fig.2, the temperature variation of $M_s$ obeys rather accurately the power law $M_s(T)=M_s(0)(1-T/T_C)^{\beta}$ with $M_s(0)=5.1~kOe$. As the most striking result we note that the exponent $\beta=0.66\pm0.02$ is not only significantly greater than $\beta=0.35$ for the 3D Heisenberg ferromagnet \cite{Sta71} but also larger than the mean field value $\beta=0.5$. We conjecture here an effect of the fluctuating dipolar fields in the disordered granular structure, but cannot yet offer a more detailed discussion of this interesting observation.

Basically, fluctuation effects should be more pronounced in the initial susceptibility, the temperature dependence of which was determined more accurately by ac-measurements and the zero-field cooled (ZFC) magnetization, see Fig.3(a). Upon decreasing temperature, the polar ('out-of-plane') susceptibility reaches soon the demagnetizing limit $\chi_{\bot}=1$, whereas the parallel ('in-plane') one, $\chi_{\|}$, increases up to frequency dependent peak temperatures $T_f$. The enveloppe on the high-temperature side of the peak defines the static limit $\chi_{\|}(T)=\chi_{\|}(T,\omega\to 0)$, which is analyzed by a Kouvel-Fisher plot \cite{KF64} in the inset. The appearance of a straight line for $\left[d (\ln \chi_{\|}) /dT\right]^{-1}$ signals a power law, $\chi_{\|}(T)=\chi_{\|}(0)(1-T/T_C)^{\kappa}$, the Curie temperature $T_C$ = 323.5 K and exponent $\kappa=2.6\pm0.1$ of which are given by the intercept with the T-axis and by the slope, respectively. Similar as for $M_s(T)$, the large exponent $\kappa$ characterizes a rather slow increase of the susceptibility, as compared e.g. to the superparamagnetic limit, $\chi_{\|}(T \to T_C)\sim\mu_g^2(T)\sim(1-T/T_C)^{2\beta}$, with $2\beta=0.70$ expected for non-interacting granular moments $\mu_g(T)=M_s(T)V_g$, where $V_g$ is the mean grain volume. This feature indicates a strong suppression of magnetic dipole fluctuations by the intergranular dipolar coupling.

\begin{figure}
\begin{center}
\includegraphics[width=8cm]{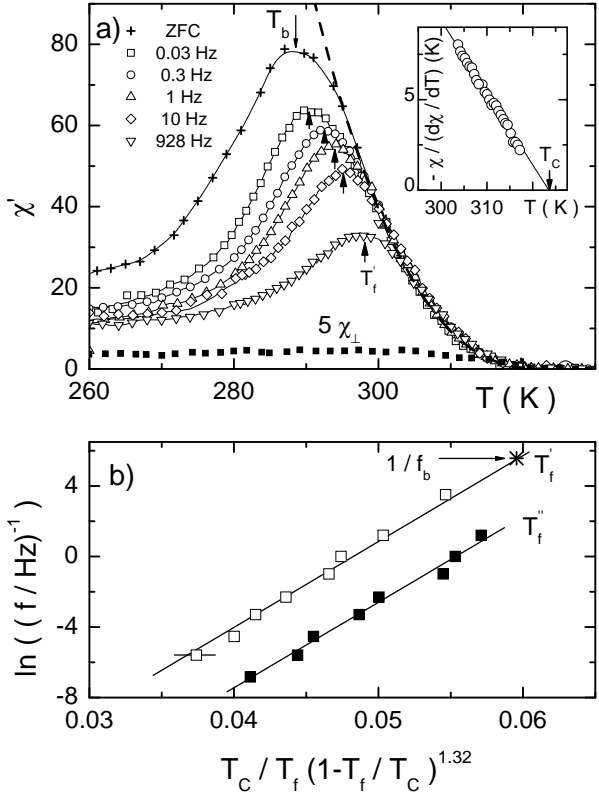}
\caption{(a) Temperature variation of the real part of the ac-susceptibility $\chi'_{||}$ of NiMnIn measured in zero magnetic field between 0.03~Hz and 928~Hz for ac-fields parallel to the film-plane. Also included are the zero-field-cooled in-plane susceptibility, $M_{ZFC}/H$, recorded during warming in H=1~Oe and the out-of-plane susceptibility $\chi'_{\bot}$ for 10~Hz. All solid curves are drawn as guides to the eye. Inset: Kouvel-Fisher analysis of the static in-plane susceptibility, $\chi_{||}(T)=\chi'_{||}(T,\omega\rightarrow0)$. The straight line defines the Curie-temperature $T_C$ and the exponent $\kappa$=2.6(1) of the power law $\chi_{||}(T)\sim(1-T/T_C)^{\kappa}$, depicted by the broken curve in the main frame. (b) Arrhenius analyses of the peak temperatures T$_f$ of $\chi'_{||}$  ( $\square$ marked by arrows in panel a) and $\chi''_{||}$ ($\blacksquare$ in Fig. 4) for a temperature dependent energy barrier, $E_b(T)=E_b(0)(1-T/T_C)^{2\beta}$.}
\end{center}
\end{figure}
 
In Fig.3(b), the freezing temperatures  $T'_{f}$ and $T''_{f}$ following from the peaks of the real (Fig.3(a)) and imaginary (Fig.4) parts of $\chi_ {||}(T,\omega=2\pi f)$, respectively, are analyzed in terms of the traditional N\'eel-Arrhenius plot for a thermally activated relaxation time of the magnetization, $\tau(T)=\tau_0 \exp(E_b(T)/k_BT)$. Here the usual 1/T variable has been modified by taking into account a temperature dependence of the energy barrier which is demanded by the fact that otherwise a completely unphysical attempt frequency $\tau^{-1}_{0}=10^{146}~s^{-1}$ is obtained. We argue that the barrier is related to the dominant anisotropy in the granular film, i.e. to the dipole-dipole interaction between the granular moments, because, as mentioned previously, the crystalline anisotropy field of the Heusler alloy is much smaller than the spontaneous magnetization $M_s$. Hence, we assume $E_b(T) \sim \mu_g^2(T)$, for which the temperature variation of $M_s(T)=\mu_g(T)/ V_g$ in Fig.2 implies $E_b(T)=E_b(0)(1-T/T_C)^{2\beta}$. Within the Debye model, 

\begin{equation}
\chi_{||}(\omega,T)=\chi_{||}(T) \frac {1}{1+i\omega\tau(T)},  
\end{equation}\\

\noindent
the peak temperatures of $\chi'$ and $\chi''$ obey the relation $E_b(T_f)/T_f=\ln(f_{0}/f)$. Hence, the slopes of both plots in Fig. 3(b) determine the barrier at zero temperature, $E_b(0)/k_B=150\cdot10^{3} K$, while the displacement of both lines arises from the difference $f'_{0}/f''_0\approx|\ln\omega\tau_0|\approx 3.5$. The preexponential time $\tau_0=1.4\cdot10^{-13}$~s follows directly from the extrapolation of the $\chi''$-peaks to $1/f''_{0} \cong 2\pi\tau_0\exp(1/|\ln\omega\tau_0|)\cong 2\pi\tau_0$ at $1/T_f=0$. We have outlined this technical point in some detail, because our time dependent MC simulations of the problem, to be presented in section III, provide only the real part of the dynamic susceptibility.

The present discussion of the dynamics in terms of a thermally activated process is also supported by the ocurrence of a blocking of the ZFC magnetization at $T_b$ in Fig. 3(a). Depicting $T_b$ on the Arrhenius plot for $T'_{f}$ in Fig. 3(b), a blocking time $\tau_b=(2\pi f_b)^{-1}=24$~s is obtained, a magnitude which is usually accepted for the onset of magnetic irreversibility \cite{BL59}. A detectable coercive field, $H_c=2$~Oe, appears near 270~K, i.e. below the temperature range of interest in this work.

\begin{figure}
\begin{center}
\includegraphics[width=8cm]{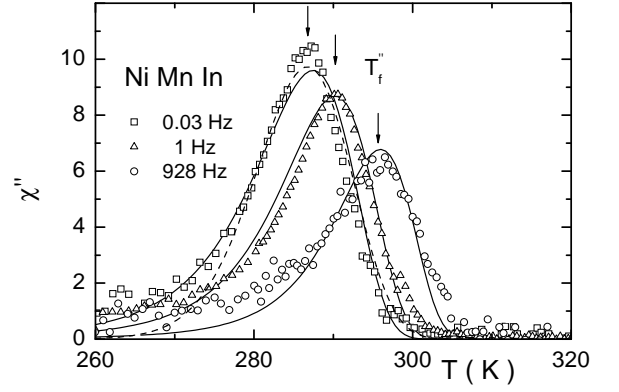}
\caption{Temperature variation of the ac-absorption of the NiMnIn film. The solid curves have been calculated from Eq.2, using a gaussian distribution of energy barriers of width $\sigma_b$=0.24 and the mean barrier $E_b(T)$ following from Fig. 3(b). The dashed curve has been calculated using the log-normal distribution depicted in Fig. 1(b).}
\end{center}
\end{figure}

The temperature dependence of the absorption component of $\chi_{\|}$ is exemplified for three frequencies in Fig. 4. As the width of the peak turns out to be much larger than for the pure Debye-shape, Eq.1, a distribution of energy barriers in the disordered granular structure is very likely. Based on the results in Fig. 1(b), we assume a gaussian distribution of granular blocking volumes $V$ about a mean value $V_g$, $P(v)=(2\pi\sigma_v)^{-\frac{1}{2}}\exp(-(v-1)^{2}/2\sigma_v)$, where $v=V/V_g$. This implies for the barrier $E_b(v)=v \cdot E_b$, where $E_b$ is the mean barrier of the grains. The ac-susceptibility computed with $\tau(T,v)=\tau_0\exp(vE_b/k_BT)$ and $\sigma_v=0.24$ from the analysis in Fig. 1(b),

\begin{equation}
\chi_{||}(\omega,T;\sigma_v)=\int_{-\infty}^{\infty}dv P(v) v~\chi_{||}(\omega,T,v)
\end{equation}\\

\noindent
agrees excellently with the data in Fig. 4. Hence, the results are fully consistent with the gaussian distribution. Alternatively, inserting the log-normal distribution from Fig. 1(b) to Eq.(2), we find also good agreement with the data, indicated by the dashed curve in Fig. 4, exept in the wing on the low temperatur side of the $\chi''$-peak. 

These findings permit the estimates of two important quantities, i.e. the mean magnetostatic energy (in SI units $\mu_0=4\pi\cdot10^{-7}$~Vs/Am) of a single grain, $E_g=\frac {1}{2}\mu_0~ M_s^2V_g$ with the mean volume of a grain $V_g=A_gt=(23.8\pm1.5)~10^3$~nm$^3$, and of the mean dipolar coupling energy between adjacent grains, 

\begin{equation}
	G_d(T)= \frac {\mu_0}{4\pi} \frac{\mu_g^2(T)}{d_g^3}.
\end{equation}\\

\noindent
At first, we compare the magnetostatic energy at T=0, $E_g/k_B=192 \cdot 10^3$ K with the energy barrier at T=0 determined from the Arrhenius analysis, $E_b(T=0)/k_B=150 \cdot 10^3$ K, to obtain $E_b=0.78~E_g$. This result suggests a close relation between the barrier and the magnetostatic energy, which for an out-of-plane flip of a granular moment sets an upper limit for $E_b$. Of course, this upper limit may not be reached due to a reduction of the in-plane dipolar field by the disorder and/or by thermal fluctuations of the adjacent moments.

\section{Monte Carlo simulations}

In order to explore the intriguing question to what extent the disorder and thermal fluctuations affect the magnetization statics and dynamics, we performed extensive MC simulations for $100\times 100$ classical spins with dipolar interaction on a triangular lattice. The lattice sites have been randomized so that gaussian distributions of granular distances $d_g$ with various widths $\sigma_d$ were realized and, hence, the central features of the granular structure of the film were taken into account.\cite{MC} 

The interaction Hamiltonian reads

\begin{equation}\label{dip}
E_{d}({\bf r}_{ij})  = \frac{1}{2}G_d \sum_{i,j} \left( \frac {{\bf S}_i\cdot
{\bf S}_j}{\widetilde{r}_{ij}^{~3}}-3 \frac{ \left({\bf S}_i\cdot {\bf r}_{ij}\right)
        \left({ \bf S}_j \cdot {\bf r}_{ij} \right) } {\widetilde{r}^{~5}_{ij}}\right),\\
\end{equation}
\noindent
where $\widetilde{r}_{ij}=r_{ij}/d_g$ is the mean distance between the moments at ${\bf r}_i$ and ${\bf r}_j$ normalized to the nearest neighbor spacing $d_g$, and $G_d$ denotes the strength of the coupling, given in Eq. 3. Our aim is to give a reasonable theoretical description of finite arrays. For that reason and in order to avoid symmetry adapted structures, open boundary conditions have been used. To prevent artificial effects we used no cutoff. A standard MC technique was used \cite{VOG00}.
In contrast to MC schemes for exchange coupled systems, where only the restricted movements
of magnetic moment are often used \cite{NCK00}, the rotational space was sampled uniformly , i.e. a moment can try any new angle. This is especially important in dipolar systems as the dipolar interaction often favors large angles between neighboring spins. An extremely slow annealing procedure with up to 150 temperature steps has been applied. The number of MC steps (MCS) at each temperature, i.e. the MC observation time $t_{MC}$,  has been varied between $0.5\cdot 10^3$ and $10^4$, so that the whole procedure contained up to $1.5 \cdot 10^6$ MCS. To avoid metastable states we have performed two different simulations of the same system simultaneously, starting them at different seeds for the random number generator to ensure that the samples take different paths to the equilibrium. Only when both samples reached the same stable energy level it has been accepted that the system has reached equilibrium.

\begin{figure}
\begin{center}
\includegraphics[width=8cm]{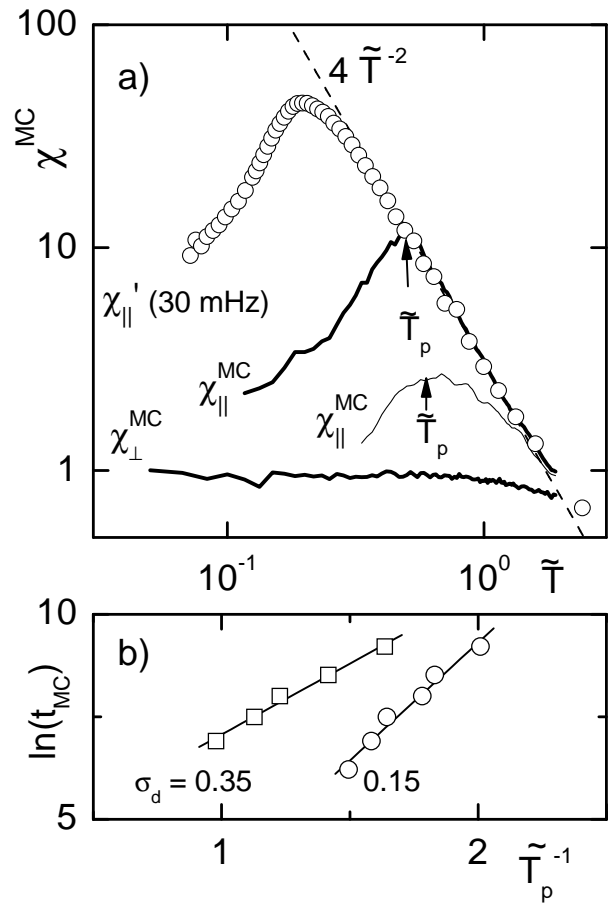}
\caption{(a) In- and out-of-plane susceptibilities $\chi_{||}$ and $\chi_{\bot}$ evaluated from Monte Carlo (MC) simulations for dipolar coupled moments  \textit{versus} temperature normalized to the  coupling constant, $\widetilde{T}=k_BT/G_d$. The MC results are obtained after $t_{MC}=5\cdot 10^3$ MC steps for 100$\times$100 classical spins on site disordered triangular lattices with gaussian widths $\sigma_d=0.15$ and 0.35 (thin curve). Above the peak temperature $\widetilde{T}_p$, the log-log plot reveals the power law $\chi_{\|}^{MC}(\widetilde{T}>\widetilde{T}_p)=4\widetilde{T}^{-2}$ (broken line), which is in absolute agreement with the experimental data for $\chi_{||}$ at 30~mHz inferred from Fig.3(a) using $G_d(T)=G_d(0)(1-T/T_C)^{1.32}$. (b) Arrhenius plot of the peak temperatures $\widetilde{T}_p$ \textit{vs}. MC time $t_{MC}$  yielding the energy barrier $E_b^{MC}$ for two different widths $\sigma_d$.}
\end{center}
\end{figure}

The central results for the in-plane and perpendicular susceptibilities, $\chi^{MC}_{\alpha}=4\pi G_d(0)/k_BT\cdot(\left\langle S_{\alpha}^2\right\rangle-\left\langle S_{\alpha}\right\rangle^2)$, $\alpha=\|$ and $\bot$, are depicted in Fig. 5(a) versus the normalized temperature $\widetilde{T}=k_BT/G_d$. Very similar to the experimental observation in Fig. 3(a), $\chi_{\|}$ displays a peak at some temperature $\widetilde{T}_p$, the position and width of which depend on the strength of the disorder. On the high temperature side of the peak, the double logarithmic plot clearly reveals a power law, $\chi^{MC}_{\|}(\widetilde{T}>\widetilde{T}_p)= 4 ~\widetilde{T}^{-2}$. A variation of the MC observation time $t_{MC}$ does not change this behavior, so that this power law represents to the equilibrium susceptibility and, hence, a direct comparison with the experimental results for $\chi_{\|}(T)$ can be made. For this purpose, we employ our $\chi'$-data for $\omega/2\pi$=30~mHz from Fig. 3(a) and transform the temperature scale to the reduced MC-scale by means of Eq.(3), $\widetilde{T}=k_BT/G_d(0)(1-T/T_C)^{1.32}$, where $G_d(0)=\mu_0M_s^2(0)V_g^2/4\pi d_g^3=33.1 \cdot 10^3~k_B$~K. Using this relation between  $\widetilde{T}$ and $T$, we see immediately that the MC-result $\chi^{MC}_{\|}(\widetilde{T}>\widetilde{T}_p)= 4 \widetilde{T}^{-2}$ implies directly $\chi_{\|}(T)= \chi(0) (1-T/T_C)^{2.64}$. As a matter of fact, $\chi_{\|}^{MC}(\widetilde{T}>\widetilde{T}_p)$ agrees not only with the T-dependence of the static suceptibility in Fig. 3(a), but we find also perfect agreement for the amplitude $\chi_{\|}(0)$ and for $\chi_{\bot}(T)$.

Before discussing the origin behind this suppression of the magnetic fluctuations in the susceptibility by help of the MC spin configurations, we examine the maximum of $\chi_{\|}^{MC}$, i.e. freezing behavior occuring at $\widetilde{T}_p$. To this end, we calculate $\chi_{\|}^{MC}(\widetilde{T})$ for MC times between $t_{MC}=5\cdot 10^2$ and $10^4$ MCS and find the peak temperature to decrease for longer $t_{MC}$. The slopes of the Arrhenius plots for $\widetilde{T}_p^{-1}$ presented in Fig. 5(b) define the normalized energy barriers $d(\ln t_{MC})/d\widetilde{T}_p^{-1}=E_b^{MC}/G_d$, which turn out to be reduced by increasing disorder $\sigma_d$. In order to compare these results
with the experimental barrier, we note that $\sigma_d$ measures the variance of the intergranular distances, so that the corresponding variance of areas is given by $\sigma_A=2\sigma_d$. Hence, the MC barrier for $\sigma_d=0.15$ may be compared to $E_b=(192\pm 10)~10^3 k_b$~K of the Heusler film with $\sigma_b=0.24$  extracted from the $\chi''(T,\omega)$-shape in Fig. 4, with which the width of the distribution function $P(v)$ in Fig. 1(b) was consistent. In fact the MC barrier for this disorder, $E_b^{MC}(0)/k_B=5.8\cdot G_d(0)/k_B=191\cdot 10^3$~K, is almost identical with the measured value. However, regarding the slight difference between the disorder strengths this identity appears to be accidental.

Motivated by the overall agreement between the experimental and the MC susceptibilities in the equilibrium and blocking regimes and for a deeper understanding of the observations it is rather suggestive to explore also the MC-generated spin configurations. Since the dipole interaction between the  moments prefers the formation of vortices \cite{VGL99}, we introduce the socalled vorticity of the triangular lattice cell at ${\bf r}_i$,

\begin{equation}
 Q_z({\bf r}_i) = \frac{d_g}{2}\cdot rot_z {\bf S}({\bf r}_i).
\end{equation}
\noindent 
Since at the temperatures of interest here the $S_z$-fluctuations are negligible, the other components can be ignored, $Q_x = Q_y = 0$.\\

At first, let us discuss the global, i.e. average of the vorticity modulus of the disc,

\begin{equation}
	Q_z=N^{-2} \sum_{i=1}^{N^2}|Q_z({\bf r}_i)|. 
\end{equation}\\

\noindent
For vanishing disorder, the groundstate consists of a single vortex or antivortex localized at the center, $Q_z({\bf r}=0)=\pm1$ ( see Fig. 6(c)), hence $Q_z=N^{-2}$ turns out to be very small. At finite $\sigma_d$, $E_d$ favors non-collinear spin configurations in regions with enhanced disorder. At low temperatures this is evidenced in  Fig. 6(a) by the strong increase of $Q_z$ with rising $\sigma_d$. For the weakest disorder, the average vorticity $Q_z$ displays at $\widetilde{T}_v=0.5$ a rather discontinuous transition upon increasing temperature. Within the accuracy of the data, this transition to a state with a maximum vorticity, $Q_z=1$, does not depend on the disorder.

\begin{figure}
\begin{center}
\includegraphics[width=8cm]{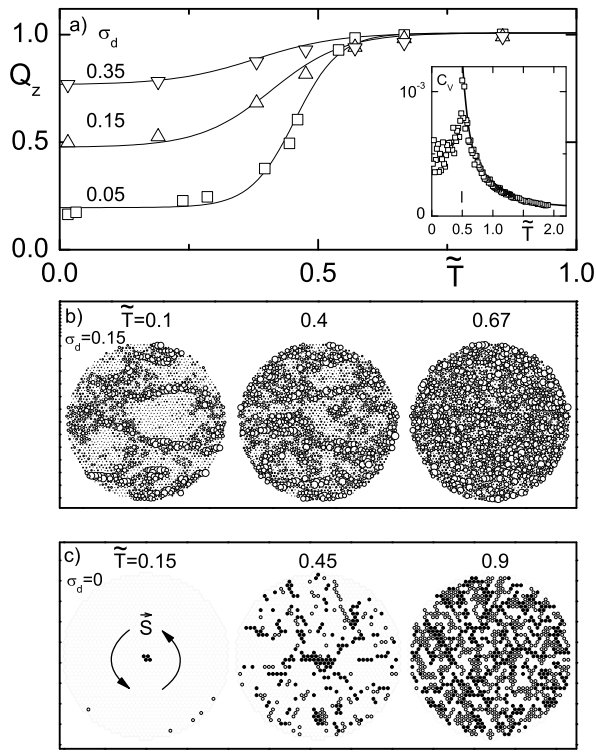}
\caption{(a) Temperature variations of the vorticity modulus, Eq. 6, indicating a thermal transition at $\widetilde{T}_v=0.5$ for vanishing disorder $\sigma_d$. This feature is supported by the divergence of the specific heat shown by the inset for $\sigma_d=0.05$. 
Vorticity maps determined from MC spin configurations at temperatures in the fluctuating $(\widetilde{T}\approx \widetilde{T}_p)$ and in the frozen state $(\widetilde{T}\ll \widetilde{T}_p)$ for gaussian disorder strengths, (b) $\sigma_{d}$=0.15, close to the experimental value, and  (c) $\sigma_d=0$. The diameter of the circles increases with the modulus of $Q_z(\vec{r}_i)$. The bright regions in panel (c) are free of vortices and display a spin circulation {\bf S} indicated for $\widetilde{T}=0.15$.}
\end{center}
\end{figure}

More specifically, the presence of a thermal transition is born out by the normalized specific heat $C_V =(\left \langle E_d^2 \right \rangle - \left \langle E_d \right \rangle^2)/(k_BT)^2$, see inset to Fig. 6(a) for $\sigma_d=0.05$. Approaching $\widetilde{T}_v$ from above, a power law singularity is found, $C_V(T)\sim (\widetilde{T}-\widetilde{T}_v)^{-\alpha}$, with $\alpha=1.73\pm0.02$,  which is independent of $t_{MC}$. This exponent is very large as compared to values for isotropic 3D ferromagnets, $\alpha\leq 0.1$ \cite{Sta71}. Taking also the slow increase of the thermal susceptibility into account, it is suggestive to attribute this behavior to the vortex fluctuations at high temperatures. As the disorder increases, the peak of $C_V(\widetilde{T})$ is observed to remain near $\widetilde{T}_v=0.5$, but being strongly suppressed. This indicates that the disorder reduces the thermal fluctuations by freezing vortices at sites which are favored by the interplay between the structural disorder and the dipolar interaction.

These effects of temperature and disorder on the vorticity $Q_z$ and the specific heat can be further illustrated by vorticity maps depicted in Fig. 6(b) for an intermediate disorder, $\sigma_d=0.15$, i.e. close to the experimental value. At low temperatures, a network of local vortex fluctuations is realized, the structure of which reflects the variation of the disorder assumed for the MC simulations.
A more detailed inspection of the spin structure and the underlying lattice disorder reveals, that the nucleation of vortices is preferred at sites where the lattice spacing is reduced and that the dipolar interaction favors a lining-up of vortices of the same sign. Such linear structures are clearly seen in Fig. 6(b) at the lower temperatures. As they separate vortex-free domains with almost homogeneous magnetizations pointing in reverse directions, these 'vortex-lines' act like domain walls in ferromagnets, and it is plausible to attribute the observation of shorter relaxation times, i.e. smaller energy barriers, at larger disorder $\sigma_d$ (see Fig.5(b)) to the presence of more walls. At the highest temperature, the entire disc is continuously covered by vortex fluctuations, which reduce further the relaxation time so that the equilibrium susceptibility, $\chi_{\|}(\widetilde{T})=4~\widetilde{T}^{-2}$, is reached in both, the experiment (Fig. 3(a)) and the MC simulation (Fig. 5(a)). On the other hand, since these dipolar-induced vortex fluctuations are non-collinear spin structures, the susceptibility is strongly suppressed as compared to the superparamagnetic limit.

Additional physical insight is provided by Fig. 6(c), where in the absence of disorder the thermal excitation of vortices is realized. Far below the transition at $\widetilde{T}_v$, i.e. in the groundstate, there is a single vortex in the center of the disc determining the direction of the circulation of the almost homogeneous magnetization $\widetilde{{\bf M}}\sim {\bf S}$ in the remainder of the sample. Just below $\widetilde{T}_v$, at $\widetilde{T}=0.45$, the number of vortices is much smaller than in the presence of disorder (Fig. 6(b)). At the larger temperature, $\widetilde{T}=0.9$, the number of vortices increases dramatically resembling a gas of thermally excited vortices. We believe that an explanation of the vortex transition at $\widetilde{T}_v=0.5$ should take into account the short range order of lining-up the vortices which appears at $\widetilde{T}=0.45$ in Fig. 6(c). On this basis, a Berezinskii-Kosterlitz-Thouless transition \cite{KT73a} as conjectured by Prakash and Henley \cite{PH90} for 2D dipolar-coupled spins seems to be unlikely. We do not see any evidence for a binding of vortices and antivortices, but rather the trend of an attraction between vortices of one kind to form linear arrays. 

\section{Summary and conclusions}

Upon decreasing the temperature below $T_C$ of a thin Heusler alloy film consisting of exchange-decoupled ferromagnetic nanograins, $V_g\approx$ (29~nm)$^3$, the effect of the strongly increasing dipolar interaction between the grains on the ordering of the film has been investigated. The rather slow rise of the zero-field susceptibility, $\chi_{\|}(T)\sim \mu_g^4(T) \sim G_d^2(T)$, is in quantitative agreement with MC simulations of dipolar-coupled classical spins on a triangular lattice. We should emphasize, that no finite size effect was observed when reducing the system from $100\times 100$ to $60\times 60$ spins. Regarding the comparison to the experiment, an effect of the circular sample shape, chosen for the simulations, is unlikely, since Fig. 6 reveals, that at the temperatures of interest $(\widetilde{T} >\widetilde{T}_v)$ thermal and disorder fluctuations of short range dominate the spin configurations. The MC spin configurations allow to attribute this rather striking feature to the thermal excitations of dense, small-scale vortices in the film plane, which are demanded by the dipolar coupling and which suppress the dipole fluctuations probed by the susceptibility. The temperature variation of the vorticity and of the specific heat, determined by the MC simulations in the limit of weak disorder, suggest the presence of a thermally induced phase transition  at $T_v=\frac{1}{2}G_d(T_v)/k_B$, which appears to be driven by vortex fluctuations of yet unknown structure.

The dynamic susceptibility of the granular film reveals a thermally activated magnetic relaxation across an energy barrier $E_b$. This barrier turns out to be slightly smaller than  the mean magnetostatic energy of a grain, $ E_b < 2\pi G_d$, and the time-dependent MC simulation reproduces this measured value $E_b$ also almost quantitatively. This agreement has been achieved for a gaussian site disorder on the triangular lattice with a variance close to that of the structural disorder of the film. This disorder of the film has been inferred from both the width of the measured ac-absorption and of the triangulation of the granular structure. As a matter of fact, the MC work predicts the energy barrier to be lowered by increasing disorder, which appears to be associated with the larger number of vortices generated by the disorder. Due to the dipolar interaction, vortices tend to form linear arrays, which act similar as domain walls in ferromagnets. Qualitatively, our observation is in analogy to the increase of the barrier reported for diluted 2D arrays with increasing $G_d$ \cite{LPT02}.

The present results on the effects of thermally and structure induced vortex fluctuations on the statics and dynamics of the magnetization as well as the nature of the transition near $T_v$ call for further efforts from both sides, experiment and theory. Experimentally, a direct proof of the vortices could perhaps be provided by near-field magnetic force microscopy. Another challenge could be the dynamics of the dipolar-coupled disordered grains in the presence of a magnetic field, in particular,  their hysteretic and switching behavior at low temperatures, $T \ll T_C$. Finally, we should mention the yet unexplained reduction of the increase of the spontaneous magnetization, $m_s(T) \sim (T_C-T)^{\beta}$, due to the large exponent $\beta=0.66$, where an effect of the disorder is not unlikely.

\section{Acknowledgment}
We are grateful to R. Anton, F. Schultz, G. Meier and U. Merkt (Hamburg) for support during the preparation, electrical and structural characterization of the NiMnIn-film.  E. Y. Vedmedenko gratefully acknowledges financial support from the DFG in the framework of the SFB 668.



\begin{thebibliography}{33}
\expandafter\ifx\csname natexlab\endcsname\relax\def\natexlab#1{#1}\fi
\expandafter\ifx\csname bibnamefont\endcsname\relax
  \def\bibnamefont#1{#1}\fi
\expandafter\ifx\csname bibfnamefont\endcsname\relax
  \def\bibfnamefont#1{#1}\fi
\expandafter\ifx\csname citenamefont\endcsname\relax
  \def\citenamefont#1{#1}\fi
\expandafter\ifx\csname url\endcsname\relax
  \def\url#1{\texttt{#1}}\fi
\expandafter\ifx\csname urlprefix\endcsname\relax\def\urlprefix{URL }\fi
\providecommand{\bibinfo}[2]{#2}
\providecommand{\eprint}[2][]{\url{#2}}

\bibitem[{\citenamefont{Heinrich and Bland}(2005)}]{HB05}
\bibinfo{author}{\bibfnamefont{B.}~\bibnamefont{Heinrich}} \bibnamefont{and}
  \bibinfo{author}{\bibfnamefont{J.~A.~C.} \bibnamefont{Bland}},
  \emph{\bibinfo{title}{Applications of nanomagnetism, Vol. III and IV}}
  (\bibinfo{publisher}{Springer-Verlag}, \bibinfo{address}{Berlin},
  \bibinfo{year}{2005}).

\bibitem[{\citenamefont{Neel}(1949)}]{Nee49}
\bibinfo{author}{\bibfnamefont{L.}~\bibnamefont{Neel}}, \bibinfo{journal}{Ann.
  Geophys.} \textbf{\bibinfo{volume}{5}}, \bibinfo{pages}{99}
  (\bibinfo{year}{1949}).

\bibitem[{\citenamefont{Jensen and Pastor}(2003)}]{JP03}
\bibinfo{author}{\bibfnamefont{P.~J.} \bibnamefont{Jensen}} \bibnamefont{and}
  \bibinfo{author}{\bibfnamefont{G.~M.} \bibnamefont{Pastor}},
  \bibinfo{journal}{New J. Phys.} \textbf{\bibinfo{volume}{5}},
  \bibinfo{pages}{68} (\bibinfo{year}{2003}).

\bibitem[{\citenamefont{Jensen}(2006)}]{Jen06}
\bibinfo{author}{\bibfnamefont{P.~J.} \bibnamefont{Jensen}},
  \bibinfo{journal}{Comput. Mater. Sci.} \textbf{\bibinfo{volume}{35}},
  \bibinfo{pages}{288} (\bibinfo{year}{2006}).

\bibitem[{\citenamefont{Politi and Pini}(2002)}]{PP02}
\bibinfo{author}{\bibfnamefont{P.}~\bibnamefont{Politi}} \bibnamefont{and}
  \bibinfo{author}{\bibfnamefont{G.}~\bibnamefont{Pini}},
  \bibinfo{journal}{Phys. Rev. B} \textbf{\bibinfo{volume}{66}},
  \bibinfo{pages}{214414} (\bibinfo{year}{2002});
  \bibinfo{author}{\bibfnamefont{P.}~\bibnamefont{Politi}}, 
  \bibinfo{author}{\bibfnamefont{G.}~\bibnamefont{Pini}}, \bibnamefont{and}
  \bibinfo{author}{\bibfnamefont{R.~L.}~\bibnamefont{Stamps}}
  \bibinfo{journal}{Phys. Rev. B} \textbf{\bibinfo{volume}{73}},
  \bibinfo{pages}{020405} (\bibinfo{year}{2006}).

\bibitem[{\citenamefont{Petracic}(2006)}]{Pet06}
\bibinfo{author}{\bibfnamefont{O.}~\bibnamefont{Petracic}},
  \bibinfo{journal}{J. Magn. Magn. Mat.} \textbf{\bibinfo{volume}{300}},
  \bibinfo{pages}{192} (\bibinfo{year}{2006}).

\bibitem[{\citenamefont{Garc\'{\i}a-Otero
  et~al.}(2000)\citenamefont{Garc\'{\i}a-Otero, Porto, Rivas, and
  Bunde}}]{GPR00}
\bibinfo{author}{\bibfnamefont{J.}~\bibnamefont{Garc\'{\i}a-Otero}},
  \bibinfo{author}{\bibfnamefont{M.}~\bibnamefont{Porto}},
  \bibinfo{author}{\bibfnamefont{J.}~\bibnamefont{Rivas}}, \bibnamefont{and}
  \bibinfo{author}{\bibfnamefont{A.}~\bibnamefont{Bunde}},
  \bibinfo{journal}{Phys. Rev. Lett.} \textbf{\bibinfo{volume}{84}},
  \bibinfo{pages}{167} (\bibinfo{year}{2000}).
\bibitem[{\citenamefont{Cheng et~al.}(2004)\citenamefont{Cheng, Jung, and
  Ross}}]{CJR04}
\bibinfo{author}{\bibfnamefont{J.~Y.} \bibnamefont{Cheng}},
  \bibinfo{author}{\bibfnamefont{W.}~\bibnamefont{Jung}}, \bibnamefont{and}
  \bibinfo{author}{\bibfnamefont{C.~A.} \bibnamefont{Ross}},
  \bibinfo{journal}{Phys. Rev. B} \textbf{\bibinfo{volume}{70}},
  \bibinfo{pages}{064417} (\bibinfo{year}{2004}).

\bibitem[{\citenamefont{Andersson et~al.}(1997)\citenamefont{Andersson,
  Djurberg, Jonsson, Svedlindh, and Nordblad}}]{ADJ97}
\bibinfo{author}{\bibfnamefont{J.-O.} \bibnamefont{Andersson}},
  \bibinfo{author}{\bibfnamefont{C.}~\bibnamefont{Djurberg}},
  \bibinfo{author}{\bibfnamefont{T.}~\bibnamefont{Jonsson}},
  \bibinfo{author}{\bibfnamefont{P.}~\bibnamefont{Svedlindh}},
  \bibnamefont{and} \bibinfo{author}{\bibfnamefont{P.}~\bibnamefont{Nordblad}},
  \bibinfo{journal}{Phys. Rev. B} \textbf{\bibinfo{volume}{56}},
  \bibinfo{pages}{13983} (\bibinfo{year}{1997}).

\bibitem[{\citenamefont{Kechrakos and Trohidou}(2000)}]{KT00}
\bibinfo{author}{\bibfnamefont{D.}~\bibnamefont{Kechrakos}} \bibnamefont{and}
  \bibinfo{author}{\bibfnamefont{K.~N.} \bibnamefont{Trohidou}},
  \bibinfo{journal}{Phys. Rev. B} \textbf{\bibinfo{volume}{62}},
  \bibinfo{pages}{3941} (\bibinfo{year}{2000});
  \bibinfo{author}{\bibfnamefont{O.}~\bibnamefont{Chubykalo-Fesenko}} \bibnamefont{and}
  \bibinfo{author}{\bibfnamefont{R.~W.} \bibnamefont{Chantrell}},
  \bibinfo{journal}{J. Appl. Phys.} \textbf{\bibinfo{volume}{97}},
  \bibinfo{pages}{10J315} (\bibinfo{year}{2005});
  \bibinfo{author}{\bibfnamefont{R.}~\bibnamefont{Dittrich}},
  \bibinfo{author}{\bibfnamefont{T.} \bibnamefont{Schrefl}},
  \bibinfo{author}{\bibfnamefont{D.}~\bibnamefont{Suess}},
  \bibinfo{author}{\bibfnamefont{W.} \bibnamefont{Scholz}},
  \bibinfo{author}{\bibfnamefont{H.}~\bibnamefont{Forster}} \bibnamefont{and}
  \bibinfo{author}{\bibfnamefont{J.} \bibnamefont{Fidler}},
  \bibinfo{journal}{J. Magn. Magn. Mat.} \textbf{\bibinfo{volume}{250}},
  \bibinfo{pages}{12} (\bibinfo{year}{2002}).

\bibitem[{\citenamefont{Iglesias and Labarta}(2004)}]{IL04}
\bibinfo{author}{\bibfnamefont{O.}~\bibnamefont{Iglesias}} \bibnamefont{and}
  \bibinfo{author}{\bibfnamefont{A.}~\bibnamefont{Labarta}},
  \bibinfo{journal}{Phys. Stat. Sol. A} \textbf{\bibinfo{volume}{201}},
  \bibinfo{pages}{3329} (\bibinfo{year}{2004});
  \bibinfo{journal}{Phys. Rev. B} \textbf{\bibinfo{volume}{70}},
  \bibinfo{pages}{144401} (\bibinfo{year}{2004}).
  

\bibitem[{\citenamefont{Scheinfein et~al.}(1996)\citenamefont{Scheinfein,
  Schmidt, Heim, and Hembree}}]{SSH96}
\bibinfo{author}{\bibfnamefont{M.~R.} \bibnamefont{Scheinfein}},
  \bibinfo{author}{\bibfnamefont{K.~E.} \bibnamefont{Schmidt}},
  \bibinfo{author}{\bibfnamefont{K.~R.} \bibnamefont{Heim}}, \bibnamefont{and}
  \bibinfo{author}{\bibfnamefont{G.~G.} \bibnamefont{Hembree}},
  \bibinfo{journal}{Phys. Rev. Lett.} \textbf{\bibinfo{volume}{76}},
  \bibinfo{pages}{1541} (\bibinfo{year}{1996}).

\bibitem[{\citenamefont{Russier et~al.}(2000)\citenamefont{Russier, Petit,
  Legrand, and Pileni}}]{RPL00}
\bibinfo{author}{\bibfnamefont{V.}~\bibnamefont{Russier}},
  \bibinfo{author}{\bibfnamefont{C.}~\bibnamefont{Petit}},
  \bibinfo{author}{\bibfnamefont{J.}~\bibnamefont{Legrand}}, \bibnamefont{and}
  \bibinfo{author}{\bibfnamefont{M.~P.} \bibnamefont{Pileni}},
  \bibinfo{journal}{Phys. Rev. B} \textbf{\bibinfo{volume}{62}},
  \bibinfo{pages}{3910} (\bibinfo{year}{2000}).

\bibitem[{\citenamefont{Aign et~al.}(1998)\citenamefont{Aign, Meyer, Lemerle,
  Jamet, Ferr\'e, Mathet, Chappert, Gierak, Vieu, Rousseaux et~al.}}]{AML98}
\bibinfo{author}{\bibfnamefont{T.}~\bibnamefont{Aign}},
  \bibinfo{author}{\bibfnamefont{P.}~\bibnamefont{Meyer}},
  \bibinfo{author}{\bibfnamefont{S.}~\bibnamefont{Lemerle}},
  \bibinfo{author}{\bibfnamefont{J.~P.} \bibnamefont{Jamet}},
  \bibinfo{author}{\bibfnamefont{J.}~\bibnamefont{Ferr\'e}},
  \bibinfo{author}{\bibfnamefont{V.}~\bibnamefont{Mathet}},
  \bibinfo{author}{\bibfnamefont{C.}~\bibnamefont{Chappert}},
  \bibinfo{author}{\bibfnamefont{J.}~\bibnamefont{Gierak}},
  \bibinfo{author}{\bibfnamefont{C.}~\bibnamefont{Vieu}},
  \bibinfo{author}{\bibfnamefont{F.}~\bibnamefont{Rousseaux}},
  \bibnamefont{et~al.}, \bibinfo{journal}{Phys. Rev. Lett.}
  \textbf{\bibinfo{volume}{81}}, \bibinfo{pages}{5656} (\bibinfo{year}{1998}).

\bibitem[{\citenamefont{Kayali and Saslow}(2004)}]{KS04}
\bibinfo{author}{\bibfnamefont{M.~A.} \bibnamefont{Kayali}} \bibnamefont{and}
  \bibinfo{author}{\bibfnamefont{W.~M.} \bibnamefont{Saslow}},
  \bibinfo{journal}{Phys. Rev. B} \textbf{\bibinfo{volume}{70}},
  \bibinfo{pages}{174404} (\bibinfo{year}{2004});
\bibinfo{author}{\bibfnamefont{Y.}~\bibnamefont{Takagaki}} \bibnamefont{and}
  \bibinfo{author}{\bibfnamefont{K.~H.} \bibnamefont{Ploog}},
  \bibinfo{journal}{Phys. Rev. B} \textbf{\bibinfo{volume}{71}},
  \bibinfo{pages}{184439} (\bibinfo{year}{2005});
  \bibinfo{author}{\bibfnamefont{M.~A.}~\bibnamefont{Ortigoza}},
  \bibinfo{author}{\bibfnamefont{R.~A.}~\bibnamefont{Klemm}}, \bibnamefont{and}
  \bibinfo{author}{\bibfnamefont{T.~S.}~\bibnamefont{Rahman}},
  \bibinfo{journal}{Phys. Rev. B} \textbf{\bibinfo{volume}{72}},
  \bibinfo{pages}{174416} (\bibinfo{year}{2005})
  
\bibitem[{\citenamefont{Luis et~al.}(2002)\citenamefont{Luis, Petroff, Torres,
  García, Bartolomé, Carrey, and Vaurès}}]{LPT02}
\bibinfo{author}{\bibfnamefont{F.}~\bibnamefont{Luis}},
  \bibinfo{author}{\bibfnamefont{F.}~\bibnamefont{Petroff}},
  \bibinfo{author}{\bibfnamefont{J.~M.} \bibnamefont{Torres}},
  \bibinfo{author}{\bibfnamefont{L.~M.} \bibnamefont{García}},
  \bibinfo{author}{\bibfnamefont{J.}~\bibnamefont{Bartolomé}},
  \bibinfo{author}{\bibfnamefont{J.}~\bibnamefont{Carrey}}, \bibnamefont{and}
  \bibinfo{author}{\bibfnamefont{A.}~\bibnamefont{Vaurès}},
  \bibinfo{journal}{Phys. Rev. Lett.} \textbf{\bibinfo{volume}{88}},
  \bibinfo{pages}{217205} (\bibinfo{year}{2002})
\bibitem[{\citenamefont{Telem-Shafir and Markovich}(2005)}]{TM05}
\bibinfo{author}{\bibfnamefont{T.}~\bibnamefont{Telem-Shafir}}
  \bibnamefont{and}
  \bibinfo{author}{\bibfnamefont{G.}~\bibnamefont{Markovich}},
  \bibinfo{journal}{J. Chem. Phys.} \textbf{\bibinfo{volume}{123}},
  \bibinfo{pages}{204715} (\bibinfo{year}{2005}).

\bibitem[{\citenamefont{Sasioglu et~al.}(2004)\citenamefont{Sasioglu,
  Sandratskii, and Bruno}}]{SSB04}
\bibinfo{author}{\bibfnamefont{E.}~\bibnamefont{Sasioglu}},
  \bibinfo{author}{\bibfnamefont{L.~M.} \bibnamefont{Sandratskii}},
  \bibnamefont{and} \bibinfo{author}{\bibfnamefont{P.}~\bibnamefont{Bruno}},
  \bibinfo{journal}{Phys.Rev.B} \textbf{\bibinfo{volume}{70}},
  \bibinfo{pages}{024427} (\bibinfo{year}{2004}).

\bibitem[{\citenamefont{Huang et~al.}(2004)\citenamefont{Huang, Lee, Kim, Lee,
  and Kudrayavtsev}}]{HLK04}
\bibinfo{author}{\bibfnamefont{M.~D.} \bibnamefont{Huang}},
  \bibinfo{author}{\bibfnamefont{N.~N.} \bibnamefont{Lee}},
  \bibinfo{author}{\bibfnamefont{B.~J.} \bibnamefont{Kim}},
  \bibinfo{author}{\bibfnamefont{Y.~P.} \bibnamefont{Lee}}, \bibnamefont{and}
  \bibinfo{author}{\bibfnamefont{Y.~V.} \bibnamefont{Kudryavtsev}},
  \bibinfo{journal}{IEEE Trans.Magn.} \textbf{\bibinfo{volume}{40}},
  \bibinfo{pages}{2757} (\bibinfo{year}{2004}).

\bibitem[{\citenamefont{Dong et~al.}(2001)\citenamefont{Dong, Lu, Xie, Chen,
  James, McKernan, and Palmstrom}}]{DLX01}
\bibinfo{author}{\bibfnamefont{J.~W.} \bibnamefont{Dong}},
  \bibinfo{author}{\bibfnamefont{J.}~\bibnamefont{Lu}},
  \bibinfo{author}{\bibfnamefont{J.~Q.} \bibnamefont{Xie}},
  \bibinfo{author}{\bibfnamefont{L.~C.} \bibnamefont{Chen}},
  \bibinfo{author}{\bibfnamefont{R.~D.} \bibnamefont{James}},
  \bibinfo{author}{\bibfnamefont{S.}~\bibnamefont{McKernan}}, \bibnamefont{and}
  \bibinfo{author}{\bibfnamefont{C.~J.} \bibnamefont{Palmstrom}},
  \bibinfo{journal}{Physica E} \textbf{\bibinfo{volume}{10}},
  \bibinfo{pages}{428} (\bibinfo{year}{2001}).

\bibitem[{\citenamefont{Kim et~al.}(2004)\citenamefont{Kim, Kudryavtsev, Rhee,
  Lee, and Lee}}]{KKR04}
\bibinfo{author}{\bibfnamefont{K.~W.} \bibnamefont{Kim}},
  \bibinfo{author}{\bibfnamefont{Y.~V.} \bibnamefont{Kudryavtsev}},
  \bibinfo{author}{\bibfnamefont{J.~Y.} \bibnamefont{Rhee}},
  \bibinfo{author}{\bibfnamefont{N.~N.} \bibnamefont{Lee}}, \bibnamefont{and}
  \bibinfo{author}{\bibfnamefont{Y.~P.} \bibnamefont{Lee}},
  \bibinfo{journal}{IEEE Trans.Magn.} \textbf{\bibinfo{volume}{40}},
  \bibinfo{pages}{2775} (\bibinfo{year}{2004}).
  
\bibitem[{\citenamefont{Albertini et~al.}(2004)\citenamefont{Albertini, Morellon, Algarabel, Ibarra, Pareti, Arnold, Calestani}}]{AMA01}
  \bibinfo{author}{\bibfnamefont{F.} \bibnamefont{Albertini}},
  \bibinfo{author}{\bibfnamefont{L.} \bibnamefont{Morellon}},
  \bibinfo{author}{\bibfnamefont{P.~A.} \bibnamefont{Algarabel}},
  \bibinfo{author}{\bibfnamefont{M.~R.} \bibnamefont{Ibarra}},
  \bibinfo{author}{\bibfnamefont{L.} \bibnamefont{Pareti}},
  \bibinfo{author}{\bibfnamefont{Z.} \bibnamefont{Arnold}},
  \bibinfo{author}{\bibfnamefont{G.} \bibnamefont{Calestani}},
  \bibinfo{journal}{J. Appl. Phys.} \textbf{\bibinfo{volume}{89}},
  \bibinfo{pages}{5614} (\bibinfo{year}{2001}).
  
\bibitem[{\citenamefont{Vedmedenko et~al.}(1999)\citenamefont{Vedmedenko,
  Ghazali, and Levy}}]{VGL99}
\bibinfo{author}{\bibfnamefont{E.~Y.} \bibnamefont{Vedmedenko}},
  \bibinfo{author}{\bibfnamefont{A.}~\bibnamefont{Ghazali}}, \bibnamefont{and}
  \bibinfo{author}{\bibfnamefont{J.-C.~S.} \bibnamefont{Levy}},
  \bibinfo{journal}{Phys. Rev. B} \textbf{\bibinfo{volume}{59}},
  \bibinfo{pages}{3329} (\bibinfo{year}{1999}).

\bibitem[{\citenamefont{Kurfi\ss
  et~al.}(2005{\natexlab{a}})\citenamefont{Kurfi\ss, Schultz, Anton, Meier, von
  Sawilski, and K\"otzler}}]{KSA05}
\bibinfo{author}{\bibfnamefont{M.}~\bibnamefont{Kurfi\ss}},
  \bibinfo{author}{\bibfnamefont{F.}~\bibnamefont{Schultz}},
  \bibinfo{author}{\bibfnamefont{R.}~\bibnamefont{Anton}},
  \bibinfo{author}{\bibfnamefont{G.}~\bibnamefont{Meier}},
  \bibinfo{author}{\bibfnamefont{L.}~\bibnamefont{von Sawilski}},
  \bibnamefont{and}
  \bibinfo{author}{\bibfnamefont{J.}~\bibnamefont{K\"otzler}},
  \bibinfo{journal}{J. Magn. Magn. Mat.} \textbf{\bibinfo{volume}{290-291}},
  \bibinfo{pages}{591} (\bibinfo{year}{2005}{\natexlab{a}});\bibinfo{author}{\bibfnamefont{ M.}~\bibnamefont{Kurfi\ss{}}}, Ph.D. thesis,
  \bibinfo{school}{Physics Dep. - University of Hamburg}
  (\bibinfo{year}{2005}), \bibinfo{note}{\protect{Cuvillier Verlag,
  G\"ottingen, ISBN 3-86537-423-9}}.

\bibitem[{\citenamefont{Webster}(1969)}]{Web69}
\bibinfo{author}{\bibfnamefont{P.~J.} \bibnamefont{Webster}},
  \bibinfo{journal}{Contemp. Phys.} \textbf{\bibinfo{volume}{10}},
  \bibinfo{pages}{559} (\bibinfo{year}{1969}).

\bibitem[{\citenamefont{Kurfi\ss
  et~al.}(2005{\natexlab{b}})\citenamefont{Kurfi\ss, Schultz, Anton, and
  Meier}}]{KS05}
\bibinfo{author}{\bibfnamefont{M.}~\bibnamefont{Kurfi\ss}},
  \bibinfo{author}{\bibfnamefont{F.}~\bibnamefont{Schultz}},
  \bibinfo{author}{\bibfnamefont{R.}~\bibnamefont{Anton}}, \bibnamefont{and}
  \bibinfo{author}{\bibfnamefont{G.}~\bibnamefont{Meier}}
  (\bibinfo{year}{2005}{\natexlab{b}}), \bibinfo{note}{unpublished}.
  
\bibitem[{tri()}]{tri}
\bibinfo{note}{MATHEMATICA 5.1, DiscreteMath 'ComputationalGeometry'}.

\bibitem[{FFT()}]{FFT}
\bibinfo{note}{MATHEMATICA 5.1, Built-in Function 'Fourier'}.

\bibitem[{\citenamefont{Stanley}(1971)}]{Sta71}
\bibinfo{author}{\bibfnamefont{H.~E.} \bibnamefont{Stanley}},
  \emph{\bibinfo{title}{Introduction to phase transitions and critical
  phenomena}} (\bibinfo{publisher}{Oxford University Press},
  \bibinfo{address}{London}, \bibinfo{year}{1971}).

\bibitem[{\citenamefont{Kouvel and Fisher}(1964)}]{KF64}
\bibinfo{author}{\bibfnamefont{J.~S.} \bibnamefont{Kouvel}} \bibnamefont{and}
  \bibinfo{author}{\bibfnamefont{M.~E.} \bibnamefont{Fisher}},
  \bibinfo{journal}{Phys. Rev.} \textbf{\bibinfo{volume}{136}},
  \bibinfo{pages}{A1626} (\bibinfo{year}{1964}).

\bibitem[{\citenamefont{Bean and Livingstone}(1959)}]{BL59}
\bibinfo{author}{\bibfnamefont{C.}~\bibnamefont{Bean}} \bibnamefont{and}
  \bibinfo{author}{\bibfnamefont{J.}~\bibnamefont{Livingstone}},
  \bibinfo{journal}{J. Appl. Phys.} \textbf{\bibinfo{volume}{30}},
  \bibinfo{pages}{120 S} (\bibinfo{year}{1959}).


\bibitem[{MC()}]{MC}
\bibinfo{note}{The program for the lattice randomization is available at\\ http://www.nanoscience.de/group\_r/members/vedmedenko}.


\bibitem[{\citenamefont{Vedmedenko et~al.}(2000)}]{VOG00}
\bibinfo{author}{\bibfnamefont{E.~Y.} \bibnamefont{Vedmedenko}},
  \bibinfo{author}{\bibfnamefont{H.~P.} \bibnamefont{Oepen}},
  \bibinfo{author}{\bibfnamefont{A.}~\bibnamefont{Ghazali}}, 
  \bibinfo{author}{\bibfnamefont{J.-C.~S.} \bibnamefont{Levy}}, \bibnamefont{and}
  \bibinfo{author}{\bibfnamefont{J.}~\bibnamefont{Kirschner}},
  \bibinfo{journal}{Phys. Rev. Lett.} \textbf{\bibinfo{volume}{84}},
  \bibinfo{pages}{5884} (\bibinfo{year}{2000}).
  
\bibitem[{\citenamefont{Nowak et~al.}(2000)\citenamefont{Nowak, Chantrell, and
  Kennedy}}]{NCK00}
\bibinfo{author}{\bibfnamefont{U.}~\bibnamefont{Nowak}},
  \bibinfo{author}{\bibfnamefont{R.~W.} \bibnamefont{Chantrell}},
  \bibnamefont{and} \bibinfo{author}{\bibfnamefont{E.~C.}
  \bibnamefont{Kennedy}}, \bibinfo{journal}{Phys. Rev. Lett.}
  \textbf{\bibinfo{volume}{84}}, \bibinfo{pages}{0163} (\bibinfo{year}{2000}).

  
\bibitem[{\citenamefont{Kosterlitz and Thouless}(1973)}]{KT73a}
\bibinfo{author}{\bibfnamefont{J.~M.} \bibnamefont{Kosterlitz}}
  \bibnamefont{and} \bibinfo{author}{\bibfnamefont{D.~J.}
  \bibnamefont{Thouless}}, \bibinfo{journal}{J. Phys. C}
  \textbf{\bibinfo{volume}{6}}, \bibinfo{pages}{1181} (\bibinfo{year}{1973});
\bibinfo{author}{\bibfnamefont{V.~L.} \bibnamefont{Berezinskii}},
  \bibinfo{journal}{Soviet. Phys. JETP} \textbf{\bibinfo{volume}{32}},
  \bibinfo{pages}{493} (\bibinfo{year}{1970}).

\bibitem[{\citenamefont{Prakash and Henley}(1990)}]{PH90}
\bibinfo{author}{\bibfnamefont{S.}~\bibnamefont{Prakash}} \bibnamefont{and}
  \bibinfo{author}{\bibfnamefont{C.~L.} \bibnamefont{Henley}},
  \bibinfo{journal}{Phys. Rev.} \textbf{\bibinfo{volume}{42}},
  \bibinfo{pages}{6574} (\bibinfo{year}{1990}).


\end{thebibliography}
\end{document}